\documentclass[letter]{aa}
\usepackage{graphicx}
\usepackage{natbib}
\usepackage{txfonts}
\bibpunct{(}{)}{;}{a}{}{,}

\newcommand{\eqb}{\begin{eqnarray}}
\newcommand{\eqe}{\end{eqnarray}}

\newcommand{\nelec}{n_{\rm e}}

\newcommand{\tesc}{t_{\rm esc}}
\newcommand{\tacc}{t_{\rm acc}}
\newcommand{\tcross}{t_{\rm cr}}
\newcommand{\gammamax}{\gamma_{\rm max}}
\newcommand{\gamflmx}{\gamma_{\rm fl,max}}
\newcommand{\gamflmn}{\gamma_{\rm fl,min}}

\newcommand{\qcal}{\cal{Q}}
\newcommand{\nph}{n_{\rm \gamma}}

\begin{document}

\title{On the rapid TeV flaring activity of \object{Markarian 501}}
\author{A. Mastichiadis \and K. Moraitis
}
\institute{Department of Physics, University of Athens, Panepistimiopolis, GR 15783 Zografos, Greece}
\offprints{A. Mastichiadis
}
\date{Received ... / Accepted ...}
\abstract{}{We investigate the one-zone SSC model of TeV blazars in the presence of electron acceleration. In this picture electrons reach a
maximum energy where acceleration saturates from a combination of synchrotron and inverse Compton scattering losses.}{We solve the spatially
averaged kinetic equations which describe the simultaneous evolution of particles and photons, obtaining the multi-wavelength spectrum as a
function of time.}{We apply the model to the rapid flare of \object{Mrk 501} of July 9, 2005 as this was observed by the MAGIC telescope and
obtain the relevant parameters for the pre-flare quasi steady state and the ones during the flare. We show that a hard lag flare can be obtained
with parameters which lie well within the range already accepted for this source. Especially the choice of a high value of the Doppler factor
seems to be necessary.}{}

%
%
\maketitle

\section{Introduction}

Blazars, a class of Active Galactic Nuclei, have multi-wavelength spectra showing two components: the first, which is generally believed to be
produced from electron synchrotron radiation, peaks, according to the object, from the optical up to X-ray regime; the second, which probably is
produced from inverse Compton scattering of the same relativistic electrons either on the synchrotron photons \citep{maraschi92} or on some
other, external to the source, photon population \citep{dermeretal92,sikora94}, peaks in the GeV to TeV $\gamma-$ray band. This emission is
thought to be produced within a relativistic flow pointing at us and therefore to be Doppler boosted.

Flaring activity is probably the major characteristic of these objects. This is especially true for the X-ray and TeV regimes which are produced
by radiation of electrons close to their maximum energy. Thus in the X-ray regime flares have been observed on timescales of hours to days
\citep{kataoka01} while in the TeV regime the variability can be even faster \citep{gaidos96,albert07,aharonian07}.

Recently it was reported \citep{albert07} that \object{Mrk 501} showed an episode of flaring activity in the TeV regime where hard $\gamma-$rays
lagged the soft ones by approximately four minutes. This flare combines therefore fast variability with
spectral evolution. Flares showing a hard lag have been observed
before in the X-ray regime \citetext{see, for example, \citealp{brinkmann05} and, more recently, \citealp{sato08}}, in addition to flares
showing a soft lag \citep{takahashi96}. However, such detailed spectral information was not possible to obtain so far in the $\gamma-$rays.

In general, soft lags can be interpreted as due to electron cooling. If electrons are either accelerated very fast to their maximum energy or
are injected as secondaries, then as the more energetic electrons cool faster,
they will produce first a variation in the hard photons followed
by a variation in the softer ones, exhibiting thus a soft lag \citetext{\citealp{kirima98} - hereafter KRM, \citealp{kima99},
\citealp{kusunose00}}.

Hard lags, by necessity, require some sort of particle acceleration -- for a different altogether explanation see \citet{albert08}.
If the timescale associated with it is faster than the cooling one, then
the radiation emitted from particles accelerating from low to high energies shows a hard lag. However, as the population of electrons moving to
high energies does not have enough time to cool before reaching the maximum energy, the amount of
produced radiation is small and therefore the
mechanism is rather inefficient. Thus hard lags will be more apparent close to the spectral cutoff, where, by definition, the radiative loss
timescale is comparable to the acceleration one (KRM).

In the present paper we examine the unique properties of the July 2005 flare of \object{Mrk 501} within the context of the one-zone SSC model
which has been modified as to include electron acceleration. Our goal is to determine whether the model can give reasonable fitting parameters
for both the steady state and the flaring activity. In Sect.~\ref{model} we outline the model, in Sect.~\ref{flare} we show the procedure of
fitting both the quasi steady state and the flare and we conclude in Sect.~\ref{discuss}.

\section{The Model}
\label{model}

In order to model the emission of \object{Mrk 501} we use a similar approach to \citet[][hereafter MK97]{mk97}, i.e. we solve
simultaneously two time-dependent kinetic equations for the distribution functions of electrons and photons.  The electron equation
reads \eqb\label{eq1} {\partial\nelec\over\partial t}+{\partial\over\partial\gamma} \left({\gamma\over\tacc}\nelec\right)+{\nelec\over\tesc}
+{\cal{L}}^{\mathrm{syn}}+{\cal{L}}^{\mathrm{ics}}
= Q(t)\delta(\gamma-\gamma_0)+{\qcal^{\mathrm{\gamma\gamma}}} \eqe 
while the photon equation is
\eqb\label{eq2}
\frac{\partial\nph}{\partial
t}+\frac{\nph}{\tcross}+\cal{L}^{\mathrm{\gamma\gamma}}+{\cal{L}}^{\mathrm{ssa}}={\qcal}^{\mathrm{syn}}+{\qcal}^{\mathrm{ics}}. \eqe The
operators ${\cal{L}}$  denote losses and escape from the system while ${\cal{Q}}$ denote injection and source terms. The unknown functions
$\nelec$ are $\nph$ are the differential number densities of electrons and photons respectively and the physical processes which are included in
the kinetic equations are: (1) electron synchrotron radiation and synchrotron self absorption (denoted by the superscripts "syn" and "ssa"
respectively); (2) inverse Compton scattering (both in the Thomson and Klein-Nishina regimes) ("ics"); (3) photon-photon pair production
("$\gamma\gamma$"). 

 The physical processes listed above are taken into account 
as described in \citet{mk95} (hereafter MK95), 
where a detailed description of the functional form of the above operators is given. The current version of the code uses also
some improvements over MK95. Thus 
for synchrotron radiation relativistic electrons 
emit the full photon spectrum \citep[see, e.g.][]{blum70} instead
of the delta-function approximation used in MK95. Also, for
inverse Compton scattering, while the electron cooling still
uses the technique described in MK95, the electron emissivity 
uses relation (2.48) of  \citet{blum70}.
Numerical tests have shown that this approach balances 
electron energy losses and total photon radiated power to within 90$\%$
of each other.

Furthermore, departing from the approach of MK97, we implement an acceleration term in
the electron kinetic equation which is characterized by an appropriate timescale ($\tacc$) and is accompanied by a term which describes particle
injection at some low energy (first term in RHS of Eq.~\ref{eq1} where $Q(t)$ is the rate of electrons which are injected at low 
energies
$\gamma_0m_\mathrm{e}c^2$). This modification allows us to follow particles as they accelerate from low to high energies. Note that this
approach is similar to the one taken in KRM. However, the present work differs from KRM in that we adopt here a one-zone model.

There are six parameters that are required to specify the
source in a stationary state. These include

1. The Doppler factor $\delta=[\Gamma(1-\beta \cos\theta)]^{-1}$.

2. The radius $R$ of the source (or, equivalently,  the
crossing time in the rest frame of the source $\tcross=R/c$).

3. The magnetic field strength $B$.

4. The acceleration timescale $\tacc$, which by the setup
of the problem must obey the relation $\tacc\ge\tcross$.

5. The timescale of particle escape of the system $\tesc$.

6. The rate of injected electrons $Q_0$ --
we note, however, that the solution turns
to be largely independent of the exact choice of $\gamma_0$
as long as this is not larger than 10.

As was shown in KRM this prescription (under the assumption of synchrotron radiation losses only) leads to an electron distribution function
which in steady state reads \eqb n_e(\gamma)\propto \gamma^{-2} \left({1\over\gamma}-{1\over\gammamax}\right) ^{(\tacc-\tesc)/\tesc} \eqe for
$\gamma_0\le\gamma\le\gammamax$, where $\gammamax$ is the Lorentz factor at which electron energy losses balance acceleration. 
For example, in the pure synchrotron case,
\eqb
\gammamax=(\beta_s\tacc)^{-1}
\eqe
where $\beta_s={4\over 3}\sigma_T c {B^2\over{8\pi m_ec^2}}$ with $\sigma_T$ the Thomson cross section. 
However, the present approach incorporates, in
addition to synchrotron, SSC losses which render the derivation of an analytic solution impossible due to complications arising from 
the Klein-Nishina limit.
One further  notes that
for $\tacc$ and $\tesc$ both independent of energy, as it was assumed in deriving the above solution, the electron distribution function is a
power law of index $s=-2-({\tacc-\tesc})/{\tesc}$ (as long as $\gamma<<\gammamax$). 
We note also that tests on the code in the pure synchrotron loss case
have shown that the electron distribution above $\gammamax$ does not drop
to zero abruptly, but rather, due to numerical diffusion, a very steep power law is produced.

The above description changes during a flare: Assume that the system has reached some stationary state. If this is perturbed in some way, i.e.
by injecting an increased number of particles in the acceleration mechanism for some time interval $\Delta\tau$, then a wave of fresh particles
will move to high energies. Assuming that the episode starts at some instant $t_0$, then at each time $t>t_0+\Delta\tau$ the fresh particles
will have Lorentz factors $\gamflmn(t)\le\gamma\le\gamflmx(t)$  with $\gamflmx(t)=\gamma_0e^{(t-t_0)/\tacc}$ and
$\gamflmn(t)=\gamma_0e^{(t-t_0-\Delta\tau)/\tacc}$. This relation holds as long as $\gamflmx\le\gammamax$. As the time evolving particle
distribution will have a higher amplitude than the steady state one, this will cause a flare in photons which will relax back to the pre-flare
state once particles of Lorentz factor $\gamflmn$ start becoming of order $\gammamax$. It is interesting to note that as losses do not come
solely from synchrotron radiation but from SSC as well, the flaring event could have an impact on the determination of $\gammamax$. More
specifically, the presence of extra photons during a flare in the system will increase the total electron energy losses and,
depending on the specific conditions, could cause $\gammamax$ to drop.

\section{The July 9 2005 flare}
\label{flare}

\begin{figure}
\resizebox{\hsize}{!}{\includegraphics{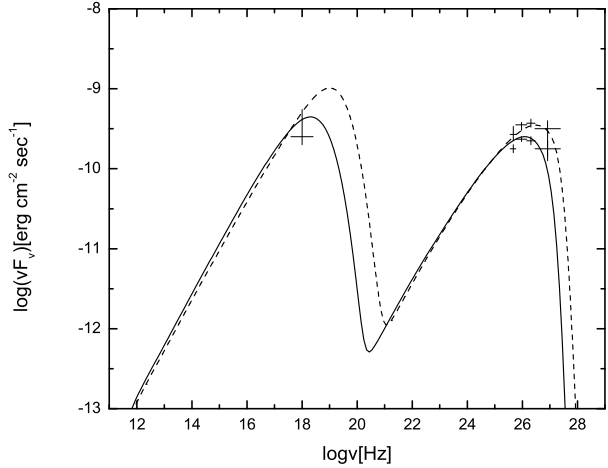}}
\caption{Fit of the multi-wavelength spectrum of \object{Mrk 501} during the flare discussed in the text. Solid curve is a fit to
an assumed steady (or pre-flare) state as discussed by \citet{albert07}. 
Parameters are as given in the text. The long dashed curve shows the
spectrum at the instance where the fluence is maximum at the hard TeV band.
The TeV spectra were de-absorbed as in
\citet{konopelkoetal03}.} \label{fig1}
\end{figure}

We begin by using the model described in the previous section to fit the X/TeV data as given in Fig.~21 of \citet{albert07} (July 9, 2005,
observations).  We have solved the set of stiff differential equations (\ref{eq1}) and (\ref{eq2}) using the numerical techniques as
these were described and tested in MK95.  The parameters used for the steady state fit are $R=1.5~10^{14}$ cm, $\delta=60$, $B=0.5$ G,
$\tacc=3\tcross$, $\tesc=4.17\tcross$, $\gamma_0=10^{0.05}$ and $Q=5~10^6$ cm$^{-3}$sec$^{-1}$erg$^{-1}$.
The cosmological parameters used are $H_0=70\ \mathrm{km}\ \mathrm{s}^{-1}\
\mathrm{Mpc}^{-1}$, $\Omega_\Lambda=0.7$ and $\Omega=0.3$. The spectrum is shown with solid line in Fig.~\ref{fig1}. The resulting electron
distribution function is a power-law of slope $s=-1.7$ up to an energy 
$\gammamax= 2~ 10^5$.

Perturbing the steady state as given above we found that a change in $Q(t)$ always produces a hard lag flare 
as the one observed.
However, this method of simulating a flaring activity causes the flux in each
energy band to increase approximately by the same amplitude -- see \citet{mamo09}.
If, as the observations seem to suggest,
the flare is becoming harder as well, i.e. more flux is emitted in the higher energy bands, then 
in order to get a fit to the TeV lightcurve one needs, in addition, to decrease $\tacc$ and/or $B$ during the flaring episode.
This can be explained from an inspection of rel. (4). A spectral hardening of the flare requires an increase
of $\gammamax$ during the episode and this can be achieved, within the context of the present
model, only by reducing one, or both, of the aforementioned parameters.  

Figures 2 and 3 depict the lightcurves resulting from such  a flaring episode.
Here the observed flare comes from an impulsive change of $Q$ by a
factor of $\sim13$ for $\Delta\tau=1t_\mathrm{cr}$ and 
a decrease of $\tacc$ and $B$ by
a factor of 1.7 for $\Delta\tau=30t_\mathrm{cr}$
which is approximately equal to the duration of the flaring episode,
i.e. to the time needed for the injected particles to reach $\gammamax$.
In order to keep the spectral slope unchanged (see rel. 3) we also 
vary $\tesc$ by the same factor.
Fig. 2 shows the lightcurve
at the lowest energy range (0.15-0.25 TeV) of the MAGIC telescope, 
while Fig. 3 shows the corresponding lightcurve at
the highest energy range (1.2-10 TeV). 
The time lag between the two energy bands
is about three minutes.

The chosen flaring episode is too short to approach the stationary state corresponding to the new increased injection. Fig.~\ref{fig1} shows,
in addition to the pre-flare steady state,
the multi-wavelength spectra at 
the instance where the fluence is maximum at the hard TeV band (long dashed line).

Fig.~\ref{fig4} shows the the derived TeV hardness ratio 
(defined as $F(1-10 \mathrm{TeV})/F(0.25-1 \mathrm{TeV})$) versus the TeV flux. 
This, as expected, is anti-clockwise.

\begin{figure}
\resizebox{\hsize}{!}{\includegraphics{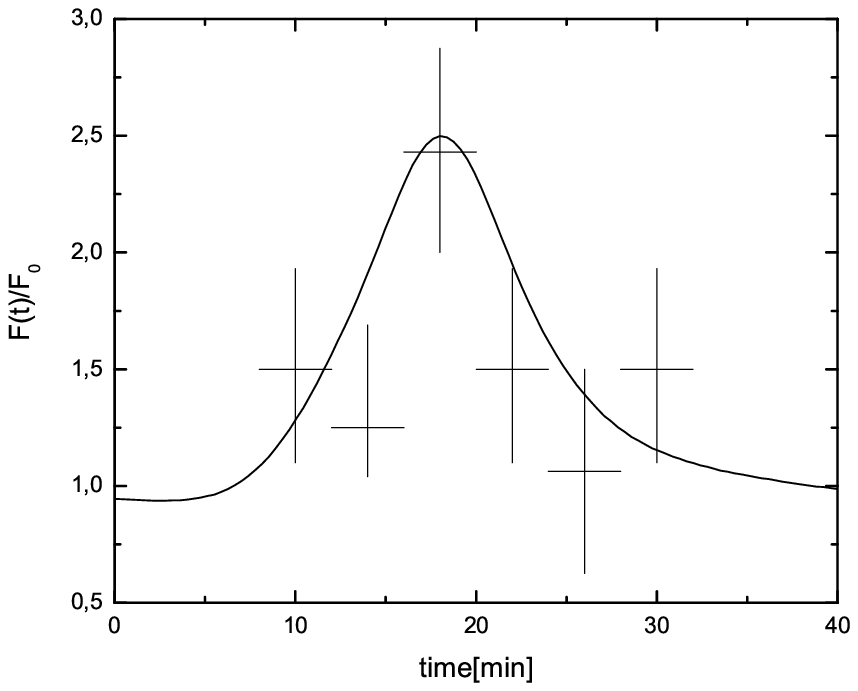}} 
\caption{Time evolution of the flare described in the text in the low (0.15-0.25 TeV) 
TeV band. Time is as measured by an observer in the lab frame. 
Observations were taken from \citet{albert07}.}
\label{fig2}
\end{figure}

\begin{figure}
\resizebox{\hsize}{!}{\includegraphics{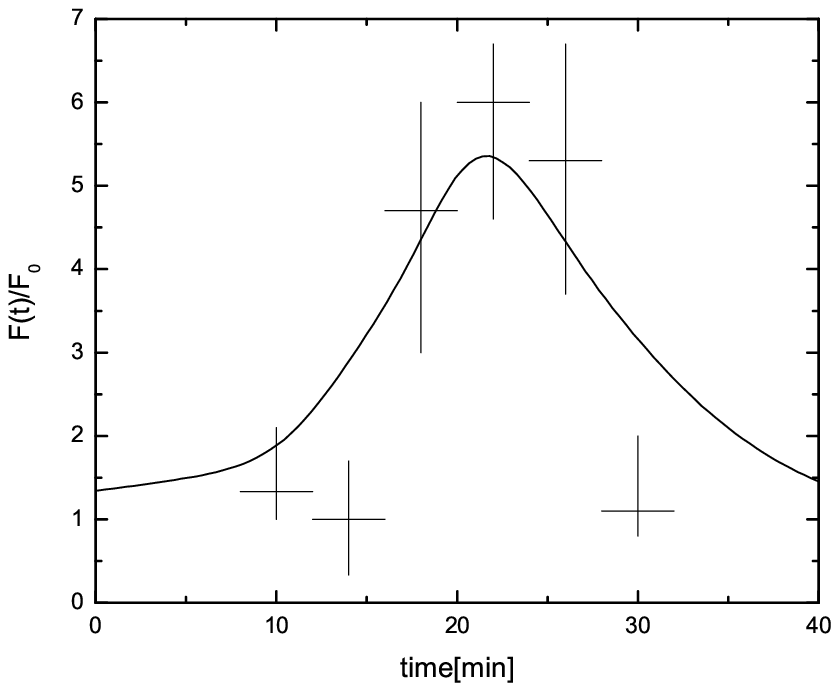}} 
\caption{Time evolution of the flare described in the text in the high (1.2-10 TeV)  TeV band. 
Time is as measured by an observer in the lab frame. 
Observations were taken from \citet{albert07}.}
\label{fig3}
\end{figure}

\begin{figure}
\resizebox{\hsize}{!}{\includegraphics{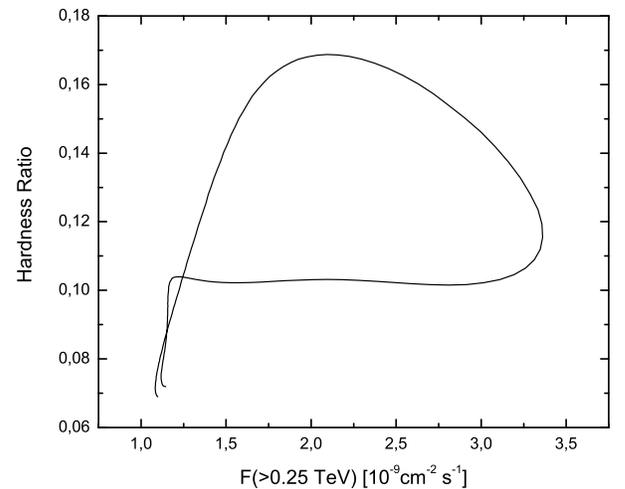}} 
\caption{Hardness ratio $F(1-10 \mathrm{TeV})/F(0.25-1 \mathrm{TeV})$ versus flux $F(>0.25\mathrm{TeV})$
for the flare described in the text. The curve rotates with time in an anti-clockwise manner.}
\label{fig4}
\end{figure}

\section{Summary/Discussion}
\label{discuss}

In the present paper we have attempted to explain the \object{Mrk 501} rapid variability as observed by the MAGIC telescope in 2005 with a
one-zone model which includes particle acceleration plus radiation. 
According to this particles are injected into some acceleration mechanism at low energies
and subsequently they  accelerated to high energies. They also radiate  synchrotron and SSC radiation, thus losing energy.
The acceleration process saturates eventually at some particle energy where it is balanced by the radiative losses.
In this picture, the basic mechanism for producing hard lags as the one observed by MAGIC
\citep{albert07}, could be a brief episode of enhanced particle injection.
As the freshly injected particles move to high energies they will radiate first soft
photons and later harder ones, thus creating a hard lag flare. This episode
occurs on top of a quasi steady state that the particles were assumed to have achieved in the
pre-flare state and to which they return in
post-flare. Thus according to this picture, MAGIC has caught electron acceleration at work.

While hard lags can be explained by increasing the particle injection at low energies, the 
observed spectral hardening of the flare requires that, in addition, the acceleration timescale and/or
the magnetic field strength should be reduced over their pre-flare values. Even if variation in one
of the above parameters can produce a spectral hardening, varying both simultaneously
improves the lightcurve fit. Given, however, the complexity of the overall flaring activity
of TeV blazars, we consider the issue to be open, requiring more detailed observations.

The parameters used here are quite similar to the ones used in \citet{konopelkoetal03}. Especially we note that in both fits there was
a need for a high value of the Doppler factor $\delta$. In the present case this choice is almost obligatory given the fast variability of the
flare \citep{fabian08}. 

The model presented here is an extension of the KRM model in the sense that it uses, in addition to synchrotron, SSC losses to balance
acceleration. However, contrary to KRM, it is an one-zone model, thus it is implicitly assumed that particles which escape do not radiate. This
can be justified only if the magnetic field strength in the region where the particles escape is much lower than in the radiating zone.
Also SSC losses are expected to be reduced away from the radiating zone as the synchrotron photon energy density drops  away from
it \citep{gould79}. Finally, we should mention that despite the fact that the electron equation (Eq.~\ref{eq1}) is similar to the one used in
the kinetic description of diffusive shock acceleration \citep[e.g.][]{ball92}, we do not specify here a detailed method for particle
acceleration. It is obvious though that simultaneous variations of more than one parameters, as is the case here,
need to be interconnected with a specific theory for acceleration. 

A prediction of this model is that a flare due to enhanced particle injection causes a
hard lag behavior in the X-ray regime as well \citep{mamo09}. However, in non-steady situations, as
it is probably the case with flares, one expects to get complicated relations between the synchrotron and the SSC component
\citep[see][]{katarz05}. 

Recently \citet{bednarek08} have proposed another explanation for the hard lag by allowing the radiating blob to accelerate during the flare.
The authors predict that, in such a case, the delay between 20 and 200 GeV should be approximately one order of magnitude longer than the
observed delay in the TeV band. On the other hand, our present model predicts that such fast flares should have the same time lag across the
$\gamma-$ray band. Thus coordinated future multi-wavelength observations involving GLAST and Cherenkov detectors could disentangle whether the
lag is due to acceleration of the blob itself or whether it is due to a brief episode of particle acceleration within a blob which moves at
constant Lorentz factor $\Gamma$.

\begin{acknowledgements}
We would like to thank the referee for comments that helped improving the paper.
This research was funded in part by a Grant from the Special Funds for Research (ELKE) of the University of Athens. KM acknowledges financial
support from the Greek State Scholarships Foundation (IKY).
\end{acknowledgements}

\bibliographystyle{aa}
\bibliography{flare}
\end{document}